\documentclass[a4paper,11pt]{article}

\usepackage{layaureo}
\usepackage[T1]{fontenc}
\usepackage[utf8]{inputenc}
\usepackage[english]{babel}
\usepackage{etex}
\usepackage{amsmath}
\usepackage{thmtools}
\usepackage{comment}
\usepackage{algorithm}
\usepackage{algorithmicx}
\usepackage{algpseudocode}
\usepackage{pgfbaseimage}
\usepackage{soul} 
\usepackage[official]{eurosym} 
\usepackage{apacite}

\usepackage{subfig}
\usepackage{tikz}
\usepackage{pgfplots}
\usepgfplotslibrary{fillbetween}
\usepgfplotslibrary{dateplot}
\pgfplotsset{compat=1.16}
\allowdisplaybreaks

\usepackage{mymacros}  

\title{Reinforcement Learning for Credit Index Option Hedging}
\author{
    Francesco Mandelli\thanks{CVA Management and A.I.~Investments, Intesa Sanpaolo IMI CIB.\\ \{francesco.mandelli, marco.pinciroli, michele.trapletti, edoardo.vittori\}@intesasanpaolo.com} \thanks{Equal contribution.}
    \and Marco Pinciroli\footnotemark[1] \footnotemark[2]
    \and Michele Trapletti\footnotemark[1] \footnotemark[2]
    \and Edoardo Vittori\footnotemark[1] \footnotemark[2]}
\date{}
\providecommand{\keywords}[1]{\small\textbf{Keywords:} #1}

\begin{document}

\maketitle

\begin{abstract}
In this paper, we focus on finding the optimal hedging strategy of a credit index option using reinforcement learning.
We take a practical approach, where the focus is on realism \textit{i.e.} discrete time, transaction costs; even testing our policy on real market data.
We apply a state of the art algorithm, the Trust Region Volatility Optimization (TRVO)~\cite{bisi2019risk} algorithm and show that the derived hedging strategy outperforms the practitioner's Black \& Scholes delta hedge. 
\end{abstract}


\section{Introduction}\label{sec:intro}

Hedging consists in investing to reduce the risk of adverse price movements of financial instruments, and it is one of the main concerns in finance. In this paper we focus on the concept of option hedging, where an option is a contract which offers the buyer the opportunity to buy or sell the underlying asset at a predefined strike price in the future. In particular, the options considered here are credit index options i.e., the underlying is a Credit Default Swap (CDS) index. Option hedging is based on a mathematical theory started with Black \& Scholes (B\&S)~\cite{black1973pricing}. This theory is motivated by a strong set of assumptions which tend to be unrealistic \cite{yalincak2012criticism}. In particular, hedging is assumed to be done costlessly and continuously. 
Several approaches have been proposed to extend the B\&S model to account for transaction costs, starting with \cite{leland1985option} and more recently \cite{gueant2017option} which uses stochastic optimal control. The main difference with respect to these approaches is that is data-driven and model-free \textit{i.e.,} it thus does not require any assumptions on the dynamics of the assets.

Credit index options market makers have the target of making profit without keeping open risk positions. The most straightforward strategy is to buy and sell the same amount of the same option in order to have a return from the difference between the two prices (the difference between the buy and sell price of a security is called bid-ask spread). Given the low liquidity of these options, most of the time this is not possible, so the market maker's portfolio results in a combination of different options, and she needs to hedge at least the risk given by the underlying instrument \textit{i.e.} the delta risk.
It is possible to hedge this risk following blindly the B\&S delta hedge, perhaps with an automated software connected directly to the market, but, specifically in cases with high transaction such as in the case of the CDS index which we are analyzing, this can quickly become quite expensive; the resulting costs may obfuscate the returns of the market maker. With other asset classes, one could concentrate on optimizing the execution thus reducing transaction costs, but in the CDS index markets this is not the case as execution costs and impact are known. So, the only way to reduce transaction costs is to minimize the transaction amount, ideally without increasing the risk related to an open delta exposure. Similar types of behavior can be found in other OTC instruments such as interest rate swaptions, so while focusing on a specific instrument, our approach remains general.
Finally, this approach is certainly interesting for XVA traders and more generically bank resource managers, who typically have to deal with portfolios with hybrid and convex risks, and experience high rebalancing costs. 

\paragraph{Contributions}
The contribution of this paper is a robust instrument capable of giving the trader a hedging signal, or even capable of autonomously trading if fitted with a market access, which is more accurate than the B\&S delta which is currently widely used, as it is optimized in discrete time and with transaction costs. Such an instrument can be created through the use of Reinforcement Learning (RL), specifically by applying TRVO~\cite{bisi2019risk}, an algorithm capable of optimizing together the hedging (i.e. risk reduction) and Profit and Loss (p\&l) objectives. By controlling the risk adverseness parameter, we are capable of creating a frontier, thus the job of the trader can be reduced to simply deciding on which point of the frontier to place himself. 
To our knowledge, this is the first time the problem of hedging credit index options is analyzed from a RL perspective and the first time this approach is tested using real data for the underlying instrument.

\paragraph{Related Works}

The issue of delta hedging using RL has been analyzed by various authors. Among the most recent approaches we mention \cite{du2020deep,kolm2019dynamic,buehler2019deep,halperin2017qlbs,halperin2019qlbs,cao19}.
These papers can be subdivided into two categories, one addresses the problem from a practitioner's perspective and is focused on the details of the hedging strategies chosen by the agent; the other builds on the formal mathematical structure of option pricing and uses machine learning techniques to overcome the problems posed by realistic features such as transaction costs.
The distinction is faint as a hedging strategy implies a price, and vice versa.

The first category includes~\cite{kolm2019dynamic,cao19} and is also pertinent for this chapter.
The most comparable, regarding the financial environment, are~\cite{du2020deep,kolm2019dynamic}, which use the same MDP formulation considered in this dissertation.
The main difference consists in the use of an approximate variance formulation in the RL objective, compared to the full variance used in this chapter. 
Furthermore,~\cite{kolm2019dynamic}, uses a one-step SARSA update, a value based approach, instead of a policy search method, while~\cite{du2020deep} considers both DQN~\cite{mnih2013playing} and PPO~\cite{schulman2017proximal}.
\cite{cao19} also consider an environment very close to ours, but with a transaction costs size that is larger than what we considered.
Regarding the RL algorithm, they use value-function methods and, in particular, risk-averse deep Q-learning. 
It is an advanced approach taken from the risk-averse reinforcement learning literature \cite{tamar2016learning}.
They consider two Q functions, one for the first moment and another for the second moment.
The paper then focuses on the agent's efficiency as a function of the rebalancing frequency. 
Differently to this chapter where we also analyze what happens when changing the risk-aversion parameter, in \cite{kolm2019dynamic, cao19}, only a single value of risk aversion is tested.

The second category includes \cite{halperin2017qlbs,halperin2019qlbs,buehler2019deep}.
In \cite{halperin2017qlbs,halperin2019qlbs}, the problem of option pricing in discrete time has been addressed from a machine learning perspective, neglecting hedging costs. 
In \cite{buehler2019deep}, the option pricing problem is undertaken by considering a class of convex risk measures and embedding them in a deep neural network environment. Initially, the dependence of the option price and hedge on the risk aversion parameter is studied in the absence of transaction costs. Then, a study of the option price dependence on transaction costs is discussed and the functional dependence of the price on the cost parameter is reconstructed. 

What distinguishes our approach is the algorithm we considered: the risk-averse policy search algorithm TRVO. One of the advantages of TRVO compared to value based algorithms like the ones used by~\cite{kolm2019dynamic,cao19,halperin2017qlbs} is the fact that being policy search, TRVO is natively compatible with continuous states and actions and thus does not suffer from the problems of using a function approximator. Furthermore, being risk-averse, it is not necessary to apply any transformation to the reward differently from what is done for example in~\cite{kolm2019dynamic} and it is able to create a policy specific on the risk aversion of the user. Moreover, an advantage of model free RL algorithms, is that the policy learned is independent from the model used to generate the data. Thus TRVO can be used as is in an option hedging framework, and only requires the standard hyperparameter tuning typical of RL algorithms.

\paragraph{Paper Outline}
In Section~\ref{sec:fin_env} we present in detail the financial framework, describing CDS indexes and options before explaining how the hedging problem can be described using Markov Decision Processes (MDPs). In Section~\ref{sec:RL} we will describe which reinforcement learning algorithm was used and for which reasons. Finally in Section~\ref{sec:experiments} we will see and evaluate the experimental performance.
\section{Financial Environment} \label{sec:fin_env}

In this section we introduce the relevant financial instruments, the Markit iTraxx Europe Senior Financial index and the credit index options built on it. We show how to price them in a standard Black \& Scholes environment and how options can be managed via standard delta hedging.

We will consider a financial environment where interest rates are set to zero for simplicity. Since both the options and the underlying are derivatives, trading them does not attract significant cash needs, and we can assume a substantial decoupling between rates and credit without loss of generality.

\subsection{Markit iTraxx Europe Senior Financial index}
The Markit iTraxx Europe Senior Financial index is a basket of credit default swaps on 30 european financial institutions (banks and insurances), equally weighted, with standardized maturities, coupons and payment dates.

Every 6 months, on the 20/09 and 20/03, or the Business Day immediately thereafter if it is not a Business Day, a new Series of the index is originated (or "rolled"). The new Series will be called "on-the-run", until a new one is generated.  Different maturities are traded for this CDS index (3, 5 and 10 years) with the maturity date that is the 20/12 or 20/06, respectively. For our purposes we consider the CDS index with 5Y maturity because it is the most liquid and there are many more options compared to the other maturities.
The index composition may be different from one Series to the other either in the number of constituents \footnote{For index versions originated before March 2015 the number of constituents was 25.} or in the CDS reference entities considered.
At the present time the Markit iTraxx Europe Senior Financial index on-the-run is the Series 35, started on 22 March 2021 and with maturity date 20 June 2026.

Each CDS index has a premium leg and a protection leg. The premium leg has standardized coupon dates: 20/03, 20/06, 20/09 and 20/12 (or the Business Day immediately thereafter if it is not a Business Day).
The coupon equals $\text{N}\,1\%\,\tau(t_{i-1},t_i) $, where N is the notional, expressed in Euro, $\tau(t_{i-1},t_i)$ is the year fraction, equal to the number of days between the present $t_i$ and the previous $t_{i-1}$ coupon date, divided by 360, while 1\% is the standardized coupon; the coupon is paid on $t_i$.\footnote{The only caveat is about the last coupon date, which corresponds to the index maturity equal to the 20/06 or 20/12 even in case that day is a holiday, with a year fraction including an extra day.}
The protection leg pays, in case of a default of one of the $j$-th Series constituents occurred before the Series' maturity, an amount equal to $\text{LGD}_j\, \text{N} \, 1/n_j$, where $\text{LGD}_j$ is the loss given default\footnote{$\text{LGD}_j$ is equal to $1-R_j$, where $R_j$ is the recovery rate determined at the end of the ISDA CDS auction triggered by the credit event.} and $1/n_j$ is the constituent weight with $n_j$ the number of constituents at the default time (at the first default $n_1=30$).
Upon default of a constituent and settlement of the relative protection leg a new version of the Series is spin-offed including the surviving constituents, the notional N is rescaled accordingly.

Since the premium leg has a standardized $1\%$ coupon, the two legs are unbalanced by an amount that is exchanged at inception as a premium, this is referred to as upfront.

Even though the upfront amount is precisely the price of the derivative, the market does not quote it directly. Rather, following the standard single name CDS convention, what is traded is the running coupon of a par (i.e. upfront equal to zero) CDS. The relation between the traded spread $S$ and the upfront, assuming the latter to be received by the protection buyer from the protection seller, is:
\begin{equation}
\label{eq:upfront}
    \text{Upf}(t,S_t) = \left(1\%-S_t\right)A_S(t)+1\% \tau(t_{acc},t),
\end{equation}
where t is the evaluation date, $\tau(t_{acc},t)$ is the year fraction, $t_{acc}$ the coupon date immediately before $t$, and $A_S(t)$ the annuity at time $t$.\footnote{In the computation of the accrual term the year fraction is modified adding an extra day.}
The latter quantity is defined as:
\begin{equation}
\label{eq:annuity}
A_S(t) = \sum_{t^+< \left\{t_i\right\} \le t_n} \tau(\max(t_{i-1}, t),t_{i}) \frac{P_S(t,t_{i-1})+P_S(t,t_i)}{2},
\end{equation}
where $t^+ = t + 1$ day, $\{t_i\}$ is the strip of index coupon dates, $t_n$ is the index maturity, $P_S(t,\theta)$ the survival probability between the present time $t$, and any future time $\theta$, given the current credit spread $S=S_t$ ($A_S(t)$ does not depend on $S_t$ directly but through $P_S(t)$).\footnote{Notice that if $t$ is the day before a coupon date, this coupon is excluded from the strip.}
The survival probability can be approximated as in~\cite{jarrow1995pricing}:
\begin{equation}
       P_S(t,\theta) = e^{-S_t\tau(t,\theta) \text{LGD}^{-1}},
\end{equation}
with LGD usually set to 60$\%$ by convention.
Making trading decisions based on the credit spread is convenient as the upfront amount has jumps at the coupon dates due to $\tau(t_{acc},t)$, while the credit spread maintains a smoother behavior.

In the following we will consider the traded spread $S$ as a sort of underlying, having it's own specific dynamics, which we will simulate with Geometric Brownian Motion (GBM). The dynamics of the index will be inherited by the dynamics of $S$.
Thus, let $S_t$ be the underlying at time $t$, then it can be described as:
\begin{equation}\label{eq:GBM}
    \mathrm{d}S_t = \mu S_t \mathrm{d}t + \sigma S_t \mathrm{d}W_t
\end{equation}
where $W_t$ is Brownian motion, $\mu$ the drift (which we assume to be 0 throughout the paper without loss of generality) and $\sigma$ the volatility.
For an initial value $S(0)$, the SDE has the analytic solution:
\begin{equation} \label{eq:spread_definition}
    S_t=S_0\exp\left(\left(\mu-\frac{\sigma^2}{2}\right)t+\sigma W_t\right)
\end{equation}
where: $W_{t+u}-W_t \sim N(0,u) = N(0,1) \sqrt{u}$.\\
\subsection{Options on the CDS index}
In this section, we consider options on the CDS index.
A {\it receiver} option gives the buyer the possibility of selling protection on the index at the expiry date at a spread equal to the strike. Conversely, a {\it payer} option gives the buyer the choice of buying protection at the expiry date at a spread equal to the strike.
Upon exercise in case of a payer (receiver) option, the option seller (buyer) physically delivers the underlying. 
In terms of the strike K and the traded spread $S(T)$ at expiry, the payoff at expiry is:
\begin{align}
\label{eq:opt_payoff_c1}
{\max \,((S_T\, A_S(T) - K\,A_K(T)), \,0)}\\
\label{eq:opt_payoff_c2}
{\max \,((K\,A_K(T)-S_T\, A_S(T)), \,0)}
\end{align}
respectively for a {\it payer} ({\it Pay}) and a {\it receiver} ({\it Rec}) option and where $ A_K(T)$ is the same expression as $ A_S(T)$ that considers $S_t=K$ in $P_S(t)$. In this paper, for simplicity we consider the payoff
\begin{align}
\label{eq:opt_payoff}
{\max \,((S_T-K)\, A_S(T), \,0)},\\
{\max \,((K-S_T)\, A_S(T), \,0)},
\end{align}
which allows a treatment \`a la Black \& Scholes on $S_t$\footnote{We focus on this simplification since the extension to the payoff of Equation~(\ref{eq:opt_payoff_c1}) and~(\ref{eq:opt_payoff_c2}), which is trivial from a numerical/RL perspective, complicates the analytical treatment in a way beyond our interest.}, since the payoff of Equation~(\ref{eq:opt_payoff}) can be seen as a call on the underlying $S_t$.

Considering an option traded at time $t$ with expiry $T$ and strike $K$, where for ease of notation we may write $S_t$ instead of $S(t)$:
$K$:
\begin{align}
\label{eq:opt_pricing payer}
& \text{Pay}(t,S_t) =\left[\Phi(d_t)S_t(T) - \Phi(e_t)K\right] A_S(T), \\
&\text{Rec}(t,S_t)=\left[\Phi(-e_t)K- \Phi(-d_t)S_t(T)\right] A_S(T), \label{eq:opt_pricing receiver}\\
& d_t = \frac{1}{\sigma\sqrt{{\tau}(t,T)}}\left[\log\left(\frac{S_t(T)}{K}\right)+\left(\frac{\sigma^2}{2}\right){\tau}(t,T)\right], \nonumber\\
&e_t =d_t-\sigma\sqrt{{\tau}(t,T)}, \nonumber
\end{align}
where $S_T(t)$ is the forward value of $S_t$, $\sigma$ is the volatility and $\bar{\tau}(t,T)$, the number of days between $t$ and $T$ divided by 365\footnote{ACT/365 convention.}\footnote{$T$ for the annuity is the settlement date, $T$ for $d_1$, $d_2$ is the expiry date.}.

Of course, as is common for options, Equation~(\ref{eq:opt_pricing payer}) and Equation~(\ref{eq:opt_pricing receiver}) are a way of mapping the option price into a volatility surface, which is convenient since the latter is a smoother function of the expiries and the strikes than the price is.\\
When trading a payer option, the buyer pays the option premium upfront to the seller, which delivers the underlying in case of exercise. There are no extra payments.
In case a name in the index defaults between $t$ and $T$, the option doesn't {\it knock out} {i.e. there is not an automatic close out} and, if exercised, it delivers {\it both} the protection leg of the defaulted name and the spin-offed index. In this sense the underlying of the option remains unchanged even if a default occurs, so that the relation between the underlying and the option is default-neutral. 
Hence, one can neglect, as we will do, jump-to-default effects in modeling the underlying and option dynamics.

Finally, since the buyer of a payer (receiver) index option receives (pays) protection substantially from trading time $t$ and not from expiry $T$, the option price needs to be adjusted consequently in order to consider any losses due to the default before $T$. This is done by a proper adjustment to the forward spread. 
Assuming zero interest rates for simplicity, the adjusted forward $S_T(t)$ is
\begin{align}
    \label{eq:opt_fwd_adj_definition_1}
S_t(T) & = S_t +  \text{LGD}(1-P_S(t,T))\frac{1}{A_S(T)}.
\end{align}
In the limit  $S(t)\rightarrow 0$, assuming the option is traded at time $t$
\begin{align}
    \label{eq:opt_fwd_adj_definition_2}
S_T(t) & \sim  S_t\left(1+\frac{\tau(t,T)}{\tau(T,t_n)}\right).
\end{align}

\subsection{Trading the Index}
\label{ssec:trading}


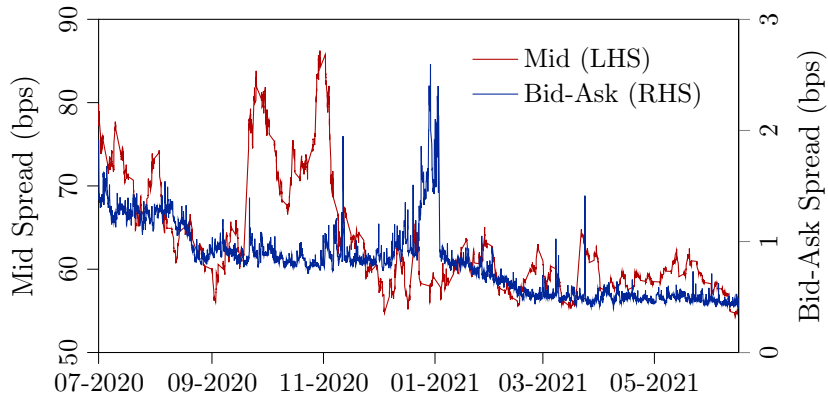
\begin{figure}[t]
    \centering
    \begin{tikzpicture}

    \pgfkeys{/pgf/number format/.cd,1000 sep={}}
    \pgfplotsset{every tick label/.append style={font=\small}}

    \begin{axis}
        [
            height=6cm, width=10cm,
            tick align=outside,
            tick pos=left,
            xmin=2020-07-01 08:30, xmax=2021-06-16 11:00,
            ymin=50, ymax=90,
            tick style={color=black},
            ylabel={Mid Spread (bps)},
            yticklabel style={rotate=90},
            date coordinates in=x,
            table/col sep=comma,
            xtick = {
                2020-07-01 08:30,
                2020-09-01 08:30,
                2020-11-01 08:30,
                2021-01-01 08:30,
                2021-03-01 08:30,
                2021-05-01 08:30
            },
            xticklabel=\month-\year,
            xticklabel style={rotate=0, anchor=near xticklabel}
        ]

        \addplot+[no markers, thin, txtRed] table[x=timestamp, y=mid] {./figures/fig_1_data.csv};
        \label{plot_mid}
    \end{axis}
    
    \begin{axis}
        [
            scaled y ticks=false,
            tick label style={
                /pgf/number format/fixed,
                /pgf/number format/fixed zerofill,
                /pgf/number format/precision=0
            },
            height=6cm, width=10cm,
            tick align=outside,
            tick pos=right,
            axis y line*=right,
            axis x line=none,
            xmin=0, xmax=4120,
            ymin=0, ymax=3,
            legend style={at={(0.57, 0.95)}, anchor=north west, draw=white},
            legend cell align={left},
            ylabel={Bid-Ask Spread (bps)},
            yticklabel style={rotate=90},
            table/col sep=comma,
        ]

        \addlegendimage{/pgfplots/refstyle=plot_mid}\addlegendentry{\small Mid (LHS)}
        \addplot+[no markers, thin, txtBlue] table[x=index, y=bidask] {./figures/fig_1_data.csv};
        \addlegendentry{\small Bid-Ask (RHS)}
    \end{axis}

\end{tikzpicture}
    \caption{The evolution of the iTraxx Europe Financial Senior 5y mid spread (in basis points, left axis) and bid/ask (in basis points, right axis) from mid 2020 to mid 2021.}
    \label{fig:prices}
\end{figure}

The iTraxx Europe Senior Financial index is traded on the so-called "over-the-counter" (OTC) market. One of the most important differences with regulated exchanges is the trade execution; in regulated markets anonymous orders are sorted and matched through an order book managed directly by the exchange. In OTC markets, CDS indexes like the Senior Financial are traded against dealers through a Multilateral Trade Facility (MTF). Dealers contribute continuously a bid and an ask price for a given notional, ranging from 10 to up to 400 mln Eur, with most of the contributions ranging between 50 and 200 mln Eur. Trading times are not strictly regulated but generally go between 9am CET and 18pm CET.
Each contributor has a typical bid/ask span, which depends on the market conditions and on the level of the index spread (a wide spread is typically correlated with a larger bid-ask).
The market spread published by the dealers can be applied, but acceptance from the dealer is not always ensured: some dealers ensure that the spread level will be always confirmed, some retain the right to review it. Moreover, dealers can also decide to publish bid and ask spreads which cannot be executed at all, and are often off market. This is another difference from the market makers' quotes in the regulated markets, which are binding. Another consequence is that it is difficult to define an order book for OTC traded objects, to distinguish if a quote is applicable or not and what would have been the maximum size executable. 

\paragraph{Defining trading costs}
We approach the trading costs problem from a statistical perspective. The starting point is a dataset containing the most recent bid and ask spread quoted by all the dealers (about 20 in the dataset) every 30 minutes, during the most liquid trading hours (9:30am CET and 17:30pm CET). 
In order to use this data it is necessary to discard quotes that are most likely typos, not executable or technological problems. Thus for each time-step and for both the bid and ask we consider the mean and standard deviation of the quotes of all the dealers and discard from the set the spreads which differ from the mean by more than two standard deviations. 
Considering the processed data, we define as applicable bid the average of the remaining bid spreads, and as applicable ask the average of the remaining ask spreads. Finally we obtain the mid spread as the average of the applicable bid and applicable ask, and the bid/ask spread as the difference between them.

An alternative approach we considered to calculate trading costs was to use the median of the unfiltered bid and ask quotes. The resulting bid/ask spreads did not differ significantly from the first method, which we ultimately considered robust enough for our purpose.

In Figure~\ref{fig:prices} we show the mid and bid/ask spreads from the cleaned dataset, on a one-year time horizon, considering intra-day data with 30-minutes time-steps. In the rest of the paper, we identify the mid with the spread $S$ introduced in Equation~\ref{eq:spread_definition}. The unit of measurement of bid-ask spread and mid spread is the Basis Point (bp) \textit{i.e.} $1\text{bp} = \frac{1}{10000}$.

The way in which applicable bid and ask spreads are built ensures that notionals up to hundreds of millions of the index can be traded at that spread, so that we can discard execution-related issues, such as slippage and assume that the trading/rebalancing costs can be computed from the bid/ask shown in Figure~\ref{fig:prices}, as a linear functional of the traded notional N:
\begin{equation} \label{eq:costdef}
    c(\text{N})= \text{N} \left|\text{Upf}_t\left(S_t\pm\frac{\text{ba}}{2}\right)-\text{Upf}_t(S_t)\right|,
\end{equation}
where $ba$ is the bid/ask, Upf$(S,t)$ the index upfront at the execution time $t$, as per Equation~\ref{eq:upfront}. In Equation~\ref{eq:costdef} the $+$ ($-$) sign should be considered when buying (selling) protection.

\subsection{Embedding in a Markov Decision Process}
\label{ssec:embedding}
The hedging problem is a sequential decision problem, where the trader needs to decide at each time-step how much of the underlying instrument to trade based on information coming from the market. This sequential decision problem can be described through a Markov Decision Process (MDP).
The p\&l in one timestep of a trader long a payer option and holding $h(t)$ in the hedging portfolio is:
\begin{align}
   p\&l & = \text{Pay}_{t+1}-\text{Pay}_t   \label{eq:delta_pl}\\  &- h_t\cdot(\text{Upf}_{t+1} - \text{Upf}_t) - c(a_t - a_{t-1}) \nonumber
\end{align}
We can define the delta hedge as:
\begin{equation*}\label{eq:CDS_delta}
N_h(t) = \left(\frac{\partial P_S(t,T)}{\partial S}\right)\left(\frac{\partial \text{ Upf}(t,S_t)}{\partial S}\right)^{-1}.
\end{equation*}
The B\&S model assures that $p\&l \rightarrow 0$ when $p\rightarrow 0$, $h(t) = N_h(t)$ and there are no transaction costs ($c(N)=0$).

From now on, we consider p>0, in particular we take as a reference point 17 rebalances per day and c(N) as defined in Equation~(\ref{eq:costdef}).


We can now transition to a reinforcement learning scenario, which will be rigorously defined in the next section. We shall define this hedging environment as a sequential decision problem, specifically as a Markov Decision Problem (MDP):
\begin{itemize}
    \item the action $a_t=h_t \in [0,1]$  
    \item the state $s_t = (S_t,\text{Pay}_t,N_h(t),a_{t-1})$
    \item the reward is Equation~(\ref{eq:delta_pl})

\end{itemize}
The above formulation is similar to what is used in \cite{vittori2020option}, and is called \textit{accounting P\&L formulation} in \cite{cao19}. 
\section{Reinforcement Learning}\label{sec:RL}

In this section, we give a brief introduction of reinforcement learning, focusing on the algorithm which we based our analysis on.

A discrete-time Markov Decision Process (MDP) is defined as a tuple $\langle\Sspace,\Aspace, \mathcal{P}, \mathcal{R}, \gamma, \mu\rangle$, where $\Sspace$ is the state space, $\Aspace$ the (continuous) action space, $\mathcal{P}(\cdot|s,a)$ is a Markovian transition model that assigns to each state-action pair $(s,a)$ the probability of reaching the next state $s'$, $\mathcal{R}(s,a)$ is a bounded reward function, $\gamma\in[0,1)$ is the discount factor, and $\mu$ is the distribution of the initial state. The policy of an agent is characterized by $\pi(\cdot|s)$, which assigns to each state $s$ an action with a certain probability.\\
We consider infinite-horizon problems in which future rewards are exponentially discounted with~$\gamma$.  Following a trajectory $\tau \coloneqq (s_0, a_0, s_1, a_1, s_2, a_2, ...$), let the returns be defined as the discounted cumulative reward:
$G = \sum_{t=0}^\infty \gamma^t \mathcal{R}(s_t,a_t).$
For each state $s$ and action $a$, the action-value function is defined as:
\begin{equation}
    Q_\pi(s,a) \coloneqq \EV_{\substack{s_{t+1}\sim \mathcal{P}(\cdot|s_{t},a_{t})\\a_{t+1}\sim\pi(\cdot|s_{t+1})}}\left[\sum_{t=0}^\infty \gamma^t \mathcal{R}(s_t,a_t)|s_0 = s, a_0 = a\right], 
    \label{eq:Q_fun}
\end{equation}
The typical RL objective is to maximize the action value function, given the initial state distribution.
This objective can be maximized in two main ways, the first is by learning the action-value function for each state and action, in general using the Bellman Equation. Once the action-value function is known, the policy is: $\pi(a|s) = \mathrm{argmax}_{a} Q(s,a)$, these algorithms are called value-based \cite{Suttonbook:1998}, and are used in~\cite{kolm2019dynamic,cao19}. This approach becomes cumbersome in a hedging environment where both states and actions are (almost) continuous. There are approaches which use function approximation to interpolate the action value function, but it is necessary to discretize the state-action space loosing precision.

The other family is instead policy search methods \cite{deisenroth2013survey}, which optimize the objective by searching directly in the policy space. They can easily handle continuous actions, learn stochastic policies in partially observable, non-Markovian environments and are robust when working in datasets with large amounts of noise \cite{moody2001learning}. 
For all these reasons, we focused on policy search algorithms~\cite{peters2008reinforcement}. 
\paragraph{Risk Averse Reinforcement Learning}
\label{ssec:riskaverse}
The typical objective is to maximize the expected cumulative rewards, which in our context means maximizing the expected cumulative p\&l. But maximizing this quantity is not the correct objective for this type of problem, in fact in an ideal B\&S model, this quantity is as close as possible to zero, which translates to optimizing a risk averse objective. 
Given the great experimental results achieved in~\cite{vittori2020option}, we decided to use the Trust Region Volatility Optimization (TRVO) algorithm defined in in~\cite{bisi2019risk}.
 
 The risk averse objective is: 
$ \eta = J+\lambda \nu^2$,  where:
\begin{equation}
    \nu^2_\pi \coloneqq (1-\gamma)\EV\left[\sum_{t=0}^\infty \gamma^t \left(\mathcal{R}(s_t,a_t)-J_\pi\right)^2\right].
    \label{eq:reward_vola} 
\end{equation}
One of the interesting things of this risk metric, to which we will refer to as \textit{reward-volatility} is the it bounds the return-variance~\cite{}. We would like to bring the reader's attention to the meaning of this reward-volatility term: it is minimizing the variations between one step and the next, in contrast to the return-variance which is minimizing the variance at the end of each path. 

In this paper we aim at training agents with different risk aversions, in order to find target balances between risk (volatility) and reward. In a static environment, this can be achieved by training each agent with a specific value of $\lambda$ and algorithm will find a minimum with a specific, $\lambda$-dependent risk-reward ratio.
Instead, in an evolving environment with variable bid ask spread, a given specific value for $\lambda$ may induce different risk-reward targets, due to the fact that the terms in the risk averse objective $\nu$ will change in value even if the market conditions remain equal (\textit{i.e.} the same action will induce different transaction costs).
This is the case in our problem as we can see from Figure~\ref{fig:prices}, where the bid/ask varies significantly.
Intuitively, we can see that the dependence of the $J$ on $ba$ will be at most linear, and typically sublinear. Thus, distortions could come from a different scaling for the variance term $\nu^2$, but as will be apparent from the experiments, also in this case the distortion is sublinear, so there is no need to implement modifications/rescaling to take the issue into account.

\begin{figure}
    \centering
    \begin{tikzpicture}

    \pgfplotsset{every tick label/.append style={font=\small}}

    \begin{axis}
        [
            height=8cm, width=10cm,
            title = \textbf{Bid-Ask: 0 bps},
            title style={at={(0.05,0.05)}, anchor=west},
            tick align=outside,
            tick pos=left,
            legend style={legend columns=1, at={(0.25,0.98)}, anchor=north, draw=white},
            legend cell align={left},
            xmin=0, xmax=679,
            ymin=0, ymax=1,
            tick style={color=black},
            ylabel={Action},
            xlabel={Timestep},
            scaled y ticks=base 10:2,
            ytick scale label code/.code={},
            yticklabel=\pgfmathprintnumber{\tick}\%,
            yticklabel style={
                rotate=90,
                /pgf/number format/.cd,fixed,precision=3
            },
            table/col sep=comma,
            xticklabel style={rotate=0, anchor=near xticklabel}
        ]

        \addplot[no markers, thin, txtRed] table[x=index, y=delta_hdg] {./figures/fig_2_data.csv};
        \addlegendentry{\small Delta Hedge}
        
        
        \addplot[no markers, thin, txtGreen] table[x=index, y=10e-5] {./figures/fig_2_data.csv};
        \addlegendentry{\small $\lambda=10$}

        \addplot[no markers, thin, txtBlue] table[x=index, y=4e-5] {./figures/fig_2_data.csv};
        \addlegendentry{\small $\lambda=4$}

        \addplot[no markers, thin, txtPurple] table[x=index, y=2e-5] {./figures/fig_2_data.csv};
        \addlegendentry{\small $\lambda=2$}
    \end{axis}

\end{tikzpicture}
    \caption{The hedging strategy chosen by agents trained at different value of  the risk aversion parameter $\lambda$ is compared with the delta hedging strategy in a zero-cost environment. In the vertical axis the hedging notional as a percentage of the option notional.}
    \label{fig:agent-vs-delta_0hc}
\end{figure}
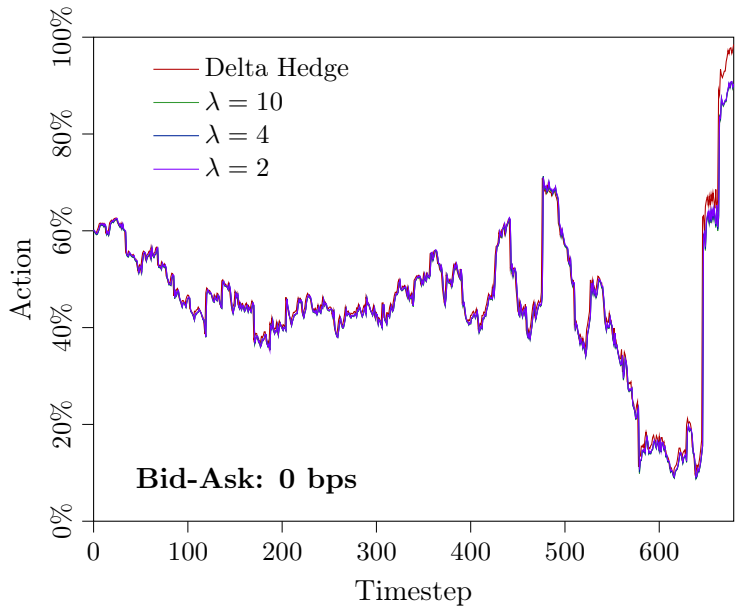

\section{Experiments}\label{sec:experiments}

In this section we present the experimental results. Once described the data generation and training parameters, we can will see the results obtained on a GBM simulated market, a heston simulated market and finally also real market data.
\subsection{Data generation and agent training}
We trained our agents on generated data, with episodes of 40 working days, with 17 observations per day, beginning at 9.30 and ending at 17.30. We simulated only the traded spread $S$, by using the GBM described in Equation~(\ref{eq:GBM}) with $\sigma$, the annualized volatility, equal to 60\% and neglecting the drift term. We did not consider the possibility of a default of one of the components as no default has been observed in recent times for the instrument in consideration.
In each simulation, the underlying spread starts from an initial value of 100 bps; we define the stochastic evolution on the actual time span between the time-steps: 30 minutes during the day, 16 hours between the last step of one trading day and the first step of the next trading day in case of two contiguous trading days, a span of $16 + 24 n$ hours in the case of trading days separated by $n$ holidays or weekend days.

We trained our agents to hedge a position short a payer (but any other position would have been equivalent) option with 2 months maturity, thus maturing at the end of each episode. The strike $K$ was 100bps, equal to the initial value of the underlying at the begging of the episode.  We assumed an option notional of 100 mln Eur, which implies an hedging portfolio containing an underlying notional between 0 and 100 mln Eur. Given the market structure, our results are valid even assuming an option notional 10 times larger . We also assumed continuous underlying trading, which is reasonable given the option size and the fact that in the market small clips (down to 100KEur or less) can be traded. Assuming a risk neutral volatility equal to 60\% the option has initial value of 530KEur.

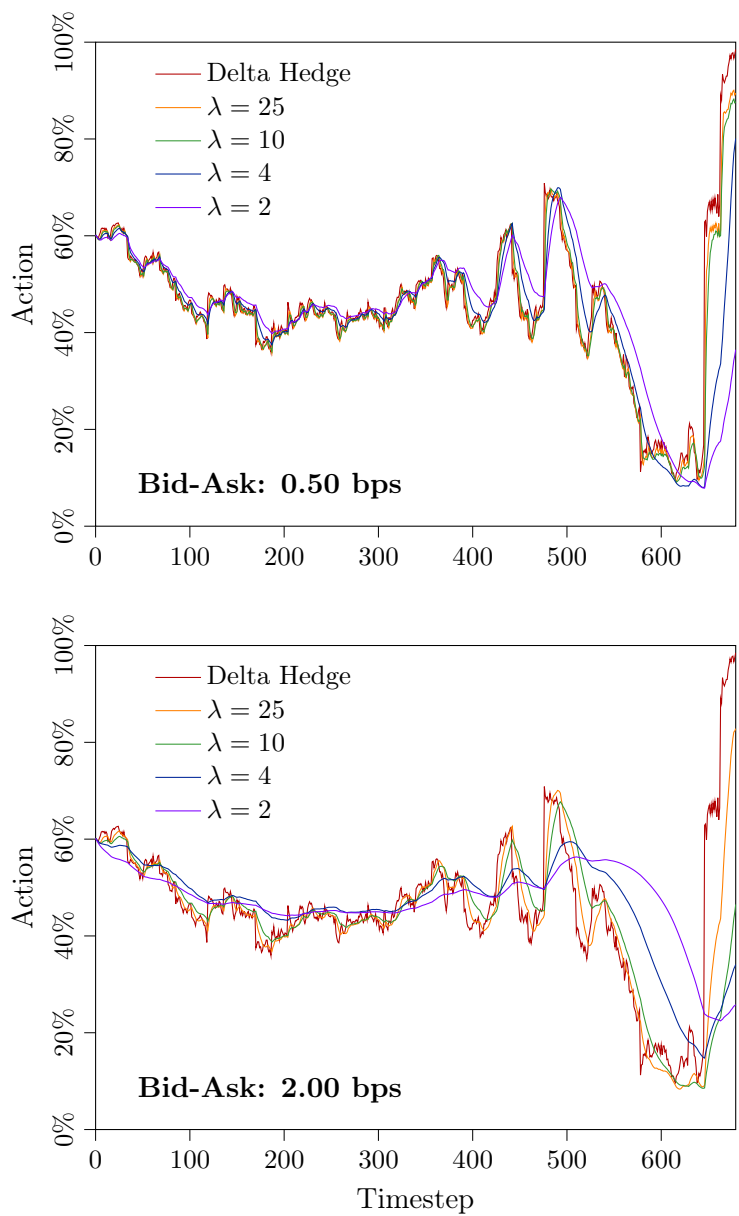
\begin{figure}
    \centering
    \begin{tikzpicture}

    \pgfkeys{/pgf/number format/.cd,1000 sep={}}
    \pgfplotsset{every tick label/.append style={font=\small}}

    \begin{axis}
        [
            height=8cm, width=10cm,
            title = \textbf{Bid-Ask: 0.50 bps},
            title style={at={(0.05,0.05)}, anchor=west},
            tick align=outside,
            tick pos=left,
            tick style={color=black},
            legend style={legend columns=1, at={(0.25,0.98)}, anchor=north, draw=white},
            legend cell align={left},
            xmin=0, xmax=679,
            ymin=0, ymax=1,
            ylabel={Action},
            table/col sep=comma,
            scaled y ticks=base 10:2,
            ytick scale label code/.code={},
            yticklabel=\pgfmathprintnumber{\tick}\%,
            yticklabel style={
                rotate=90,
                /pgf/number format/.cd,fixed,precision=3
            },
        ]

        \addplot[no markers, thin, txtRed] table[x=index, y=delta_hdg] {./figures/fig_3a_data.csv};
        \addlegendentry{\small Delta Hedge}
        
        \addplot[no markers, thin, txtOrange] table[x=index, y=25e-5] {./figures/fig_3a_data.csv};
        \addlegendentry{\small $\lambda=25$}
        
        \addplot[no markers, thin, txtGreen] table[x=index, y=10e-5] {./figures/fig_3a_data.csv};
        \addlegendentry{\small $\lambda=10$}

        \addplot[no markers, thin, txtBlue] table[x=index, y=4e-5] {./figures/fig_3a_data.csv};
        \addlegendentry{\small $\lambda=4$}

        \addplot[no markers, thin, txtPurple] table[x=index, y=2e-5] {./figures/fig_3a_data.csv};
        \addlegendentry{\small $\lambda=2$}
    \end{axis}

    \begin{axis}
        [
            yshift=-8cm,
            height=8cm, width=10cm,
            title = \textbf{Bid-Ask: 2.00 bps},
            title style={at={(0.05,0.05)}, anchor=west},
            tick align=outside,
            tick pos=left,
            tick style={color=black},
            legend style={legend columns=1, at={(0.25,0.98)}, anchor=north, draw=white},
            legend cell align={left},
            xmin=0, xmax=679,
            ymin=0, ymax=1,
            xtick style={color=black},
            ylabel={Action},
            xlabel={Timestep},
            table/col sep=comma,
            xticklabel style={rotate=0, anchor=near xticklabel},
            scaled y ticks=base 10:2,
            ytick scale label code/.code={},
            yticklabel=\pgfmathprintnumber{\tick}\%,
            yticklabel style={
                rotate=90,
                /pgf/number format/.cd,fixed,precision=3
            },
        ]

        \addplot[no markers, thin, txtRed] table[x=index, y=delta_hdg] {./figures/fig_3b_data.csv};
        \addlegendentry{\small Delta Hedge}
        
        \addplot[no markers, thin, txtOrange] table[x=index, y=25e-5] {./figures/fig_3b_data.csv};
        \addlegendentry{\small $\lambda=25$}
        
        \addplot[no markers, thin, txtGreen] table[x=index, y=10e-5] {./figures/fig_3b_data.csv};
        \addlegendentry{\small $\lambda=10$}

        \addplot[no markers, thin, txtBlue] table[x=index, y=4e-5] {./figures/fig_3b_data.csv};
        \addlegendentry{\small $\lambda=4$}

        \addplot[no markers, thin, txtPurple] table[x=index, y=2e-5] {./figures/fig_3b_data.csv};
        \addlegendentry{\small $\lambda=2$}
    \end{axis}

\end{tikzpicture}
    \caption{The hedging strategy chosen by agents trained at different value of  the risk aversion parameter $\lambda$ is compared with the delta hedging strategy in an environment including hedging costs. In the vertical axis the hedging notional as a percentage of the option notional. In the upper (lower) plot the bid/ask equals 0.5 (2) basis points.}
    \label{fig:agent-vs-delta_HC}
\end{figure}

We built a training set of 40,000 episodes, and trained our agents varying two parameters: the risk aversion parameter $\lambda$ and the bid/ask parameter $ba$. We considered $\lambda $ following~\cite{vittori2020option}, in order to span an efficient frontier in the risk/reward space; we considered $10^{-6}\lesssim\lambda\lesssim10^{-3}$, for better interpretability, in the rest of the paper we will rescale $\lambda$ by $10^5$, so to have bounds between 0.1 and 100. The choice of $ba$ as an extra parameter comes from the observation that the bid/ask of the instrument considered here shows a highly dynamic pattern (see Figure~\ref{fig:prices}).

We considered \textit{ba} ranging from 0.5 to 2 basis points (bp) as per Figure~\ref{fig:prices}. We also considered the case with low values of $ba$, even $ba=0$ in order to further test our algorithms and to check that the standard delta-hedging strategy is smoothly recovered in the limit $ba\rightarrow 0$.

\begin{figure}
    \centering
    \begin{tikzpicture}

    \pgfplotsset{every tick label/.append style={font=\small}}

    \begin{axis}
        [
            height=8cm, width=12cm,
            tick align=outside,
            tick pos=left,
            legend style={legend columns=1, at={(0.25,0.98)}, anchor=north, draw=white},
            legend cell align={left},
            xmin=40000, xmax=250000,
            ymin=0, 
            tick style={color=black},
            ylabel={$\Delta$P\&L (KEur)},
            scaled x ticks=base 10:-3,
            xtick scale label code/.code={},
            xlabel={P\&L Volatility (KEur)},
            scaled y ticks=base 10:-3,
            ytick scale label code/.code={},
            yticklabel style={
                /pgf/number format/fixed,
                rotate=90
            },
            table/col sep=comma,
            xticklabel style={rotate=0, anchor=near xticklabel}
        ]

        \draw[] (133000, 27000) node[right]{\scriptsize Delta Hedge};
        \draw[<-, thin] (67000, 2000)--(135000, 25000);
        \draw[<-, thin] (82000, 2000)--(135000, 25000);
        \draw[<-, thin] (101000, 2000)--(135000, 25000);
        \draw[<-, thin] (122000, 2000)--(135000, 25000);

        \draw[txtRed] (250000, 66000) node[left]{\small Bid-Ask: 0.50 bps};
        \addplot+[
            only marks,
            txtRed,
            mark options={solid, scale=0.5},
            mark=*,
            nodes near coords,
            point meta=explicit symbolic,
            visualization depends on={value \thisrow{Anchor_0_25}\as\myanchor},
            every node near coord/.append style={font=\scriptsize,anchor=\myanchor}
        ] table[x=Var_0_25, y=Ret_0_25, meta=Meta_0_25] {./figures/fig_4_data.csv};
        \addplot[no markers, thin, txtRed] table[x=index, y=0_25] {./figures/fig_4_interp.csv};
        \draw [dashed, txtRed] (64585, 0) -- (64585, 40000);
        
        \draw[txtOrange] (250000, 124000) node[left]{\small Bid-Ask: 1.00 bps};
        \addplot+[
            only marks,
            txtOrange,
            mark options={solid, scale=0.5},
            mark=*,
            nodes near coords,
            point meta=explicit symbolic,
            visualization depends on={value \thisrow{Anchor_0_5}\as\myanchor},
            every node near coord/.append style={font=\scriptsize,anchor=\myanchor}
        ] table[x=Var_0_5, y=Ret_0_5, meta=Meta_0_5] {./figures/fig_4_data.csv};
        \addplot[no markers, thin, txtOrange] table[x=index, y=0_5] {./figures/fig_4_interp.csv};
        \draw [dashed, txtOrange] (80144, 0) -- (80144, 94000);

        \draw[txtGreen] (250000, 183000) node[left]{\small Bid-Ask: 1.50 bps};
        \addplot+[
            only marks,
            txtGreen,
            mark options={solid, scale=0.5},
            mark=*,
            nodes near coords,
            point meta=explicit symbolic,
            visualization depends on={value \thisrow{Anchor_0_75}\as\myanchor},
            every node near coord/.append style={font=\scriptsize,anchor=\myanchor}
        ] table[x=Var_0_75, y=Ret_0_75, meta=Meta_0_75] {./figures/fig_4_data.csv};
        \addplot[no markers, thin, txtGreen] table[x=index, y=0_75] {./figures/fig_4_interp.csv};
        \draw [dashed, txtGreen] (99232, 0) -- (99232, 155000);

        \draw[txtBlue] (250000, 242000) node[left]{\small Bid-Ask: 2.00 bps};
        \addplot+[
            only marks,
            txtBlue,
            mark options={solid, scale=0.5},
            mark=*,
            nodes near coords,
            point meta=explicit symbolic,
            visualization depends on={value \thisrow{Anchor_1}\as\myanchor},
            every node near coord/.append style={font=\scriptsize,anchor=\myanchor}
        ] table[x=Var_1, y=Ret_1, meta=Meta_1] {./figures/fig_4_data.csv};
        \addplot[no markers, thin, txtBlue] table[x=index, y=1] {./figures/fig_4_interp.csv};
        \draw [dashed, txtBlue] (120179, 0) -- (120179, 217000);

    \end{axis}

\end{tikzpicture}
    \caption{Each dot represents the performance of an agent on a GBM-simulated market in terms of p\&l (w.r.t. delta hedging) and p\&l volatility, depending on $\lambda $ (annotated next to each dot) and the $ba$ parameter.}
    \label{fig:brownian_frontier}
\end{figure}
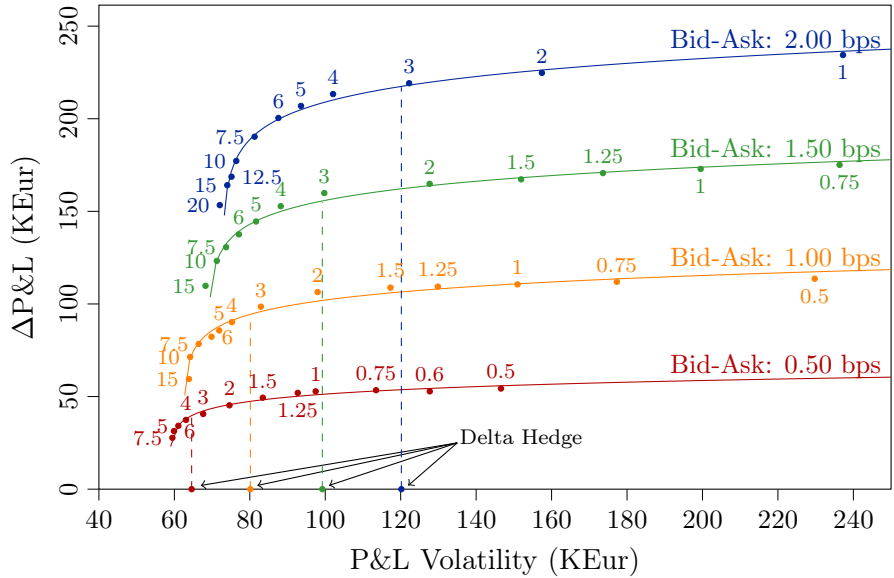

\subsection{Testing on a GBM-simulated market}
We tested our agents on a dataset of 2,000 episodes with the underlying spreads generated by the GBM, with the same parameters of the training dataset. We performed different tests varying the $ba$ spread in order to monitor agents' performances comparing to the delta-hedging strategy.

In the $ba=0$ case, the trained agent perfectly replicates the delta-hedge, this can be seen in Figure~\ref{fig:agent-vs-delta_0hc} where, for a specific testing scenario, the delta hedging strategy (in red) is compared with the action chosen by the agents trained with different values of the risk aversion parameter (in green, blue and purple). Given the absence of trading costs, all the agents replicate the same strategy, which is the optimal one\footnote{Indeed, neglecting hedging costs, the Black\&Scholes paradigm is violated only by the assumed time discretization.}, minimizing risks. Under the $ba=0$ assumption, the strategy has in average zero cumulated p\&l.

Introducing hedging costs $ba>0$ the average cumulated p\%l of the delta hedging strategy is shifted to negative values, depending linearly on $ba$, specifically, considering a $ba$ of 1 bp the cumulated p\&l is on average -136kEur. The presence of hedging costs during training induces a smoother strategy for the agent, in terms of underlying allocation changes. Since each action becomes more expensive as $ba$ increases, the agent cuts costs through the reduction of portfolio rebalances. The downside of this approach is an increase the variability of the rewards, since the option is not continuously hedged. The desired balance between cost reduction and low reward volatility can be achieved changing the lambda aversion parameter of the model. This relationship is plotted in Figure~\ref{fig:agent-vs-delta_HC}, where different degrees of smoothness in the variation of the hedging portfolio can be seen to be dependent on $\lambda$.
The smoothness degree depends also on the size of the hedging cost: defining a certain risk aversion, a higher $ba$ implies a higher smoothness, as it is apparent by comparing the upper and lower plot.

\begin{figure}
    \centering
    \begin{tikzpicture}

    \pgfplotsset{every tick label/.append style={font=\small}}

    \begin{axis}
        [
            height=7cm, width=9cm,
            tick align=outside,
            tick pos=left,
            xmin=-50000, xmax=550000,
            ymin=0, ymax=20,
            tick style={color=black},
            table/col sep=comma,
            ybar=0pt,
            bar width=6.5pt,
            area style,
            yticklabel=\pgfmathprintnumber{\tick}\%,
            yticklabel style={
                rotate=90,
                /pgf/number format/.cd,fixed,precision=3
            },
            scaled x ticks=base 10:-3,
            xtick scale label code/.code={},
            xlabel={P\&L Model - P\&L Delta Hedge (KEur)},
        ]

        \addplot[txtRed, fill=txtRed] table[x=index, y=red] {./figures/fig_5_data.csv};

        \addplot[txtGreen, fill=txtGreen] table[x=index, y=green] {./figures/fig_5_data.csv};

    \end{axis}

\end{tikzpicture}
    \caption{The distribution of the p\&l of the $\lambda=4$ agent relative to the p\&l of delta hedge assuming a the $ba$ parameter equal to 1.5 bps.}
    \label{fig:gain_dist}
\end{figure}
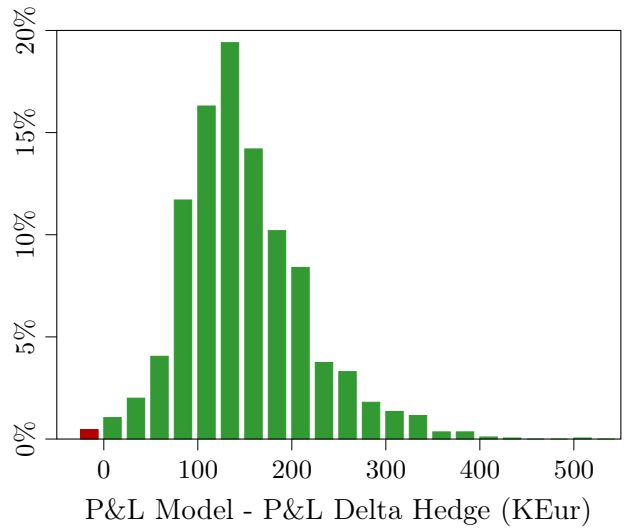

The performance of the agents w.r.t. the delta hedging strategy in terms of cumulated p\&l for different values of $\lambda$ and the $ba$ parameter is summarized in Figure~\ref{fig:brownian_frontier}.
In the figure, each dot represents the performance of an agent having the $\lambda$ indicated by the nearby annotated number and acting in an environment with $ba$ depending on the color (red for 0.5 basis point, orange for 1 basis point etc.).
The position on the vertical axis indicates the average p\&l performance of the agent w.r.t. to the delta hedging strategy in an environment having the same $ba$.
The average is taken with respect to the terminal p\&l measured on the 2,000 testing scenarios.
The position on the horizontal axis, instead, indicates the square root of the variance of the terminal p\&l (the p\&l volatility) on the same testing sample.
The colored dots laying on the horizontal axis indicate the variance performance of the delta hedging strategy in terms of p\&l volatility at different values of the $ba$ parameter.
As se can see, all the agents perform better than the corresponding delta-hedging strategies in terms of p\&l, while a certain number of agents (those lying left of the corresponding colored vertical line) perform better than the delta hedging strategy also in terms of p\&l volatility. In this sense, all the frontiers dominate the corresponding delta hedge, and it is striking to notice that the level of dominance depends on the $ba$ parameter: at low costs the dominance is mild (as it was also experienced in \cite{vittori2020option}, where the very low hedging costs of listed equity products have been considered), at high costs the delta hedging is barely reasonable a strategy. As an example, one can consider the $\lambda =4$ point of the blue frontier (which assumes very large costs and beats delta hedging both in terms of p\&l and p\&l volatility) and observe from Figure~\ref{fig:agent-vs-delta_HC} how smooth its action is.
Another thing to notice is the $\lambda$ parametrization of the different frontiers, there is a shift of $\lambda$ to the left at the increase of the $ba$ parameter. This $\lambda$-scaling in $ba$, which is very mild, is in agreement with the considerations made at the end of Section~\ref{sec:RL}.
The benefit of adopting our approach instead of the delta hedging strategy is apparent also from Figure~\ref{fig:gain_dist}, where we show the distribution of the p\&l of the $\lambda =4$ agent relative to the p\&l of delta hedging in the realistic case of $ba = 1.5$ bp: the agent essentially performs always better.

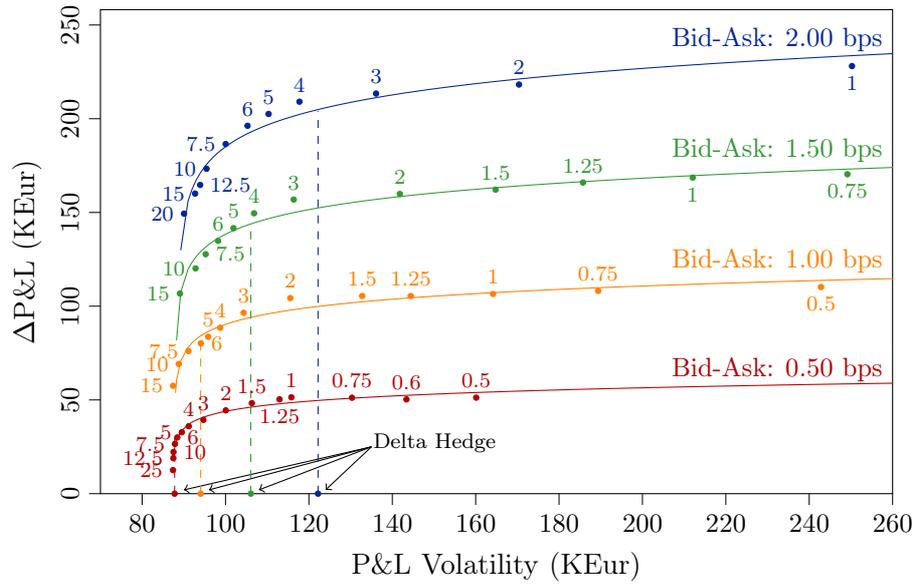
\begin{figure}
    \centering
    \begin{tikzpicture}

    \pgfplotsset{every tick label/.append style={font=\small}}

    \begin{axis}
        [
            height=8cm, width=12cm,
            tick align=outside,
            tick pos=left,
            legend style={legend columns=1, at={(0.25,0.98)}, anchor=north, draw=white},
            legend cell align={left},
            xmin=70000, xmax=260000,
            ymin=0, 
            tick style={color=black},
            ylabel={$\Delta$P\&L (KEur)},
            scaled x ticks=base 10:-3,
            xtick scale label code/.code={},
            xlabel={P\&L Volatility (KEur)},
            scaled y ticks=base 10:-3,
            ytick scale label code/.code={},
            yticklabel style={
                /pgf/number format/fixed,
                rotate=90
            },
            table/col sep=comma,
            xticklabel style={rotate=0, anchor=near xticklabel}
        ]
        
        \draw[] (133000, 27000) node[right]{\scriptsize Delta Hedge};
        \draw[<-, thin] (90000, 2000)--(135000, 25000);
        \draw[<-, thin] (96000, 2000)--(135000, 25000);
        \draw[<-, thin] (108000, 2000)--(135000, 25000);
        \draw[<-, thin] (124000, 2000)--(135000, 25000);

        \draw[txtRed] (260000, 66000) node[left]{\small Bid-Ask: 0.50 bps};
        \addplot+[
            only marks,
            txtRed,
            mark options={solid, scale=0.5},
            mark=*,
            nodes near coords,
            point meta=explicit symbolic,
            visualization depends on={value \thisrow{Anchor_0_25}\as\myanchor},
            every node near coord/.append style={font=\scriptsize,anchor=\myanchor}
        ] table[x=Var_0_25, y=Ret_0_25, meta=Meta_0_25] {./figures/fig_6_data.csv};
        \addplot[no markers, thin, txtRed] table[x=index, y=0_25] {./figures/fig_6_interp.csv};
        \draw [dashed, txtRed] (87796, 0) -- (87796, 30000);

        \draw[txtOrange] (260000, 124000) node[left]{\small Bid-Ask: 1.00 bps};
        \addplot+[
            only marks,
            txtOrange,
            mark options={solid, scale=0.5},
            mark=*,
            nodes near coords,
            point meta=explicit symbolic,
            visualization depends on={value \thisrow{Anchor_0_5}\as\myanchor},
            every node near coord/.append style={font=\scriptsize,anchor=\myanchor}
        ] table[x=Var_0_5, y=Ret_0_5, meta=Meta_0_5] {./figures/fig_6_data.csv};
        \addplot[no markers, thin, txtOrange] table[x=index, y=0_5] {./figures/fig_6_interp.csv};
        \draw [dashed, txtOrange] (94044, 0) -- (94044, 85000);

        \draw[txtGreen] (260000, 183000) node[left]{\small Bid-Ask: 1.50 bps};
        \addplot+[
            only marks,
            txtGreen,
            mark options={solid, scale=0.5},
            mark=*,
            nodes near coords,
            point meta=explicit symbolic,
            visualization depends on={value \thisrow{Anchor_0_75}\as\myanchor},
            every node near coord/.append style={font=\scriptsize,anchor=\myanchor}
        ] table[x=Var_0_75, y=Ret_0_75, meta=Meta_0_75] {./figures/fig_6_data.csv};
        \addplot[no markers, thin, txtGreen] table[x=index, y=0_75] {./figures/fig_6_interp.csv};
        \draw [dashed, txtGreen] (106074, 0) -- (106074, 144000);

        \draw[txtBlue] (260000, 242000) node[left]{\small Bid-Ask: 2.00 bps};
        \addplot+[
            only marks,
            txtBlue,
            mark options={solid, scale=0.5},
            mark=*,
            nodes near coords,
            point meta=explicit symbolic,
            visualization depends on={value \thisrow{Anchor_1}\as\myanchor},
            every node near coord/.append style={font=\scriptsize,anchor=\myanchor}
        ] table[x=Var_1, y=Ret_1, meta=Meta_1] {./figures/fig_6_data.csv};
        \addplot[no markers, thin, txtBlue] table[x=index, y=1] {./figures/fig_6_interp.csv};
        \draw [dashed, txtBlue] (122190, 0) -- (122190, 204000);

    \end{axis}

\end{tikzpicture}
    \caption{Eeach dot represents the performance on a Heston-simulated market of an agent in terms of p\&l (w.r.t. delta hedging) and p\&l volatility, depending on $\lambda $ (the number close to the dot) and the $ba$ parameter.}
    \label{fig:heston_frontier}
\end{figure}

\subsection{Testing on a Heston-simulated market}
In order to make a further step towards realism, we challenge the assumption of the GBM constant volatility, as we know it does not hold in the financial markets. We thus generated a new testing set of 2,000 episodes with spreads derived from the Heston model, which introduces a dynamic for the volatility:
\begin{align}
dS_t & = \sqrt{\nu_t}\, S_t\, dW_t^S \\
d\nu_t & = \kappa\, (\theta - \nu_t)\,dt + \xi \sqrt{\nu_t}dW^\nu_t
\end{align}
with $\nu_0 = 60\%^2$, so to recover the initial volatility used in training, $\kappa = 2$, $\theta = \nu_0$, $\xi = 0.9$, and no correlation between the stochastic terms $dW_t^S$ and $dW^\nu_t$. With this configuration $\nu_t$ oscillates significantly reaching values as high as $\sim 120\%$ and as low as $\sim 0\%$. 
When pricing the option we maintained the B\&S formulation with $\sigma=60\%$.


Even if the agents were trained on a dataset generated with GBM, they are able to achieve very good performance over the heston dataset (see Figure~\ref{fig:heston_frontier}). The reason could be that the hedging of an option is a task that implies a deep knowledge of the relationship between the underlying price and the option premium, but the way in which the underlying evolves is probably a secondary aspect. 

\begin{figure}
    \centering
    \begin{tikzpicture}

    \pgfkeys{/pgf/number format/.cd,1000 sep={}}
    \pgfplotsset{every tick label/.append style={font=\small}}

    \begin{axis}
        [
            height=8cm, width=10cm,
            tick align=outside,
            ytick pos=right,
            xtick pos=bottom,
            xmin=2020-07-21 09:30, xmax=2020-09-15 17:30,
            ymin=50, ymax=80,
            tick style={color=black},
            ylabel={S (bps)},
            yticklabel style={rotate=90},
            axis y line*=right,
            date coordinates in=x,
            table/col sep=comma,
            xtick = {
                2020-07-21 09:30,
                2020-08-04 09:30,
                2020-08-18 09:30,
                2020-09-02 09:30,
                2020-09-15 17:30
            },
            xticklabel=\day-\month-\year,
            xticklabel style={rotate=0, anchor=near xticklabel}
        ]

        \addplot+[no markers, thick, dashed, black] table[x=timestamp, y=spread] {./figures/fig_7a_data.csv};
        \label{plot_spread_a}
    \end{axis}

    \begin{axis}
        [
            height=8cm, width=10cm,
            tick align=outside,
            tick pos=left,
            tick style={color=black},
            legend style={legend columns=2, at={(0.65,0.97)}, anchor=north, draw=white},
            legend cell align={left},
            xmin=0, xmax=679,
            ymin=0, ymax=1,
            ylabel={Action},
            table/col sep=comma,
            axis x line=none,
            scaled y ticks=base 10:2,
            ytick scale label code/.code={},
            yticklabel=\pgfmathprintnumber{\tick}\%,
            yticklabel style={
                rotate=90,
                /pgf/number format/.cd,fixed,precision=3
            },
        ]
        
        \addlegendimage{/pgfplots/refstyle=plot_spread_a}\addlegendentry{\small S (RHS)}
        
        \addplot+[no markers, thin, txtRed] table[x=index, y=delta_hdg] {./figures/fig_7a_data.csv};
        \addlegendentry{\small Delta Hedge}
        
        \addplot+[no markers, thin, txtOrange] table[x=index, y=25e-5] {./figures/fig_7a_data.csv};
        \addlegendentry{\small $\lambda=25$}
        
        \addplot+[no markers, thin, txtGreen] table[x=index, y=10e-5] {./figures/fig_7a_data.csv};
        \addlegendentry{\small $\lambda=10$}

        \addplot+[no markers, thin, txtBlue] table[x=index, y=4e-5] {./figures/fig_7a_data.csv};
        \addlegendentry{\small $\lambda=4$}

        \addplot+[no markers, thin, txtPurple] table[x=index, y=2e-5] {./figures/fig_7a_data.csv};
        \addlegendentry{\small $\lambda=2$}
    \end{axis}

\end{tikzpicture}
    \caption{The delta hedging strategy (red line) is compared to the strategy selected by agents trained with $ba=1$, at various risk aversions (other colored lines), on a real episode (the black line shows the underlying spread $S$ observed between July and September 2020.}
    \label{fig:real_data_0}
\end{figure}
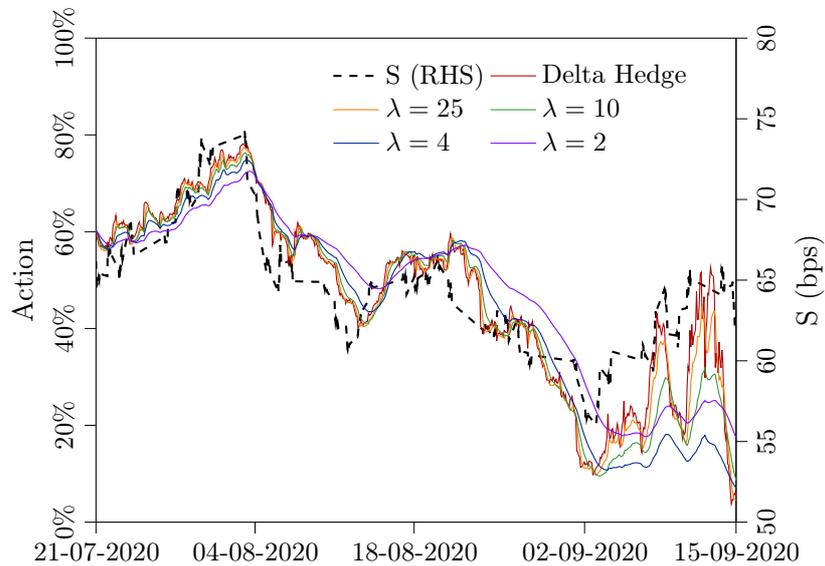
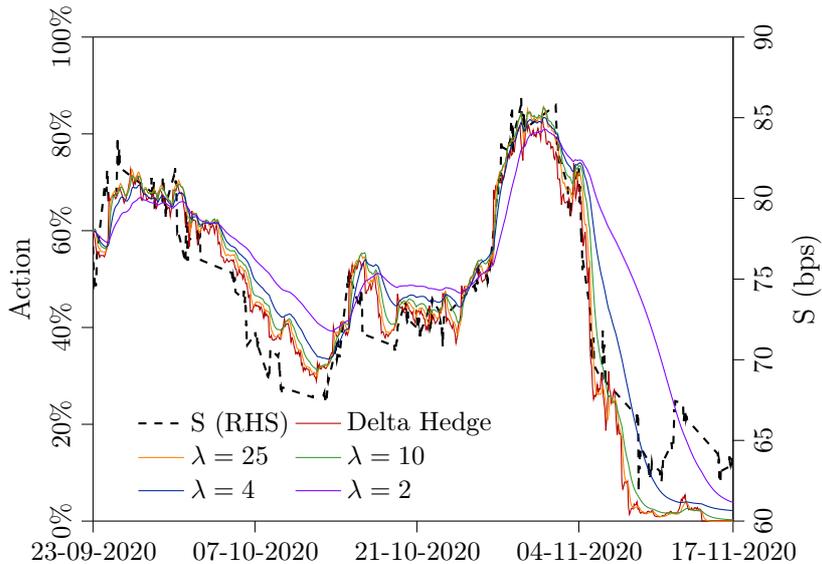
\begin{figure}
    \centering
    \begin{tikzpicture}

    \pgfkeys{/pgf/number format/.cd,1000 sep={}}
    \pgfplotsset{every tick label/.append style={font=\small}}

    \begin{axis}
        [
            height=8cm, width=10cm,
            tick align=outside,
            ytick pos=right,
            xtick pos=bottom,
            xmin=2020-09-23 09:30, xmax=2020-11-17 17:30,
            ymin=60, ymax=90,
            tick style={color=black},
            ylabel={S (bps)},
            yticklabel style={rotate=90},
            axis y line*=right,
            date coordinates in=x,
            table/col sep=comma,
            xtick = {
                2020-09-23 09:30,
                2020-10-07 09:30,
                2020-10-21 09:30,
                2020-11-04 09:30,
                2020-11-17 17:30
            },
            xticklabel=\day-\month-\year,
            xticklabel style={rotate=0, anchor=near xticklabel}
        ]

        \addplot+[no markers, thick, dashed, black] table[x=timestamp, y=spread] {./figures/fig_7b_data.csv};
        \label{plot_spread_b}
    \end{axis}

    \begin{axis}
        [
            height=8cm, width=10cm,
            tick align=outside,
            tick pos=left,
            tick style={color=black},
            legend style={legend columns=2, at={(0.35,0.25)}, anchor=north, draw=white},
            legend cell align={left},
            xmin=0, xmax=679,
            ymin=0, ymax=1,
            ylabel={Action},
            table/col sep=comma,
            axis x line=none,
            scaled y ticks=base 10:2,
            ytick scale label code/.code={},
            yticklabel=\pgfmathprintnumber{\tick}\%,
            yticklabel style={
                rotate=90,
                /pgf/number format/.cd,fixed,precision=3
            },
        ]
        
        \addlegendimage{/pgfplots/refstyle=plot_spread_b}\addlegendentry{\small S (RHS)}
        
        \addplot+[no markers, thin, txtRed] table[x=index, y=delta_hdg] {./figures/fig_7b_data.csv};
        \addlegendentry{\small Delta Hedge}
        
        \addplot+[no markers, thin, txtOrange] table[x=index, y=25e-5] {./figures/fig_7b_data.csv};
        \addlegendentry{\small $\lambda=25$}
        
        \addplot+[no markers, thin, txtGreen] table[x=index, y=10e-5] {./figures/fig_7b_data.csv};
        \addlegendentry{\small $\lambda=10$}

        \addplot+[no markers, thin, txtBlue] table[x=index, y=4e-5] {./figures/fig_7b_data.csv};
        \addlegendentry{\small $\lambda=4$}

        \addplot+[no markers, thin, txtPurple] table[x=index, y=2e-5] {./figures/fig_7b_data.csv};
        \addlegendentry{\small $\lambda=2$}
    \end{axis}

\end{tikzpicture}
    \caption{The delta hedging strategy (red line) is compared to the strategy selected by agents trained with $ba=1$, at various risk aversions (other colored lines), on a real episode (the black line shows the underlying spread $S$ observed between September and November 2020.}
    \label{fig:real_data_1}
\end{figure}

\subsection{Testing on real market data}
In order to move a further step towards a realistic setup, we consider now real market data for iTraxx Europe Senior Financial index. We use the dataset constructed in Section~\ref{ssec:trading}, thus considering real market prices and real transaction parameters $ba$ as seen in Figure~\ref{fig:prices}, and simulate the option price with $\sigma=60\%$. The available data is sufficient for 5 episodes of 40 days, which we used as a test set for agents trained with different values of $\lambda$ with $ba=1$ bp.


In Figures~\ref{fig:real_data_0} and \ref{fig:real_data_1} we show the action of the the various agents compared with the delta hedge. We also show the market data dynamics (in black), on the right vertical axis. We can see as in the previous Figures, how lower values of $lambda$ generate smoother hedging policies.

Table~\ref{tab:results} summarizes the performance of the various agents (all the figures in kEur): {\it in all the scenarios, all the considered agents 
overperform the delta hedging strategy in terms of p\&l}.  Considering risk, given the low number of scenarios at hand, the cumulated p\&l volatility previously considered is a very noisy estimator, thus, we considered the volatility of the p\&l along each scenario, a measure similar to the {\it reward volatility} defined in Equation~(\ref{eq:reward_vola}) and used in  Section~\ref{sec:RL} to define the objective of the TRVO algorithm, as described in \cite{bisi2019risk}. Using this measure no agent outperforms delta hedging, but the volatility increase is anyway very small when compared with the cost reduction obtained by adopting our agents.

\begin{table}
\small
\begin{center}
{
\begin{tabular}{| c | c | c | c | c | c | c | c | c | c | c |}
\cline{2-11}
\multicolumn{1}{c|}{} & \multicolumn{10}{c|}{Scenario}\\
\cline{2-11}
\multicolumn{1}{c|}{} & \multicolumn{2}{c|}{1} & \multicolumn{2}{c|}{2} & \multicolumn{2}{c|}{3} & \multicolumn{2}{c|}{4} & \multicolumn{2}{c|}{5} \\ \hline
$\lambda$ & p\&l & vol  & p\&l & vol  & p\&l & vol  & p\&l & vol  & p\&l & vol  \\\hline
1 &-177&4.1&183&8.7&-257&3.8&-174&2.6&-282&2.4
\\
1.25 &-165&3.6&221&8.9&-224&3.6&-185&2.4&-272&2.4
\\
1.5 &-174&3.5&236&8.9&-212&3.6&-200&2.1&-270&2.4
\\
2 &-181&3.2&219&7.1&-188&3.5&-223&1.8&-285&2.3
\\
3 &-188&3.5&154&5.3&-144&3.6&-215&1.8&-294&2.3
\\
4 &-212&3.6&95&3.9&-129&3.7&-234&1.7&-308&2.4
\\
5 &-224&3.5&62&3.3&-134&3.4&-242&1.7&-317&2.4
\\
6 &-229&3.4&40&3.0&-138&3.3&-247&1.7&-323&2.4
\\
7.5 &-231&3.0&18&2.7&-145&3.0&-252&1.6&-329&2.4
\\
10 &-237&2.8&-15&2.2&-154&2.8&-259&1.6&-338&2.4
\\
15& -264&2.4&-53&1.9&-191&2.7&-269&1.6&-349&2.3
\\
25& -293&2.2&-80&1.7&-224&2.6&-278&1.6&-356&2.3
\\\hline
$\delta$ & -376&2.1&-149&1.6&-310&2.5&-299&1.6&-372&2.3
\\ \hline
\end{tabular}}
\end{center}
\caption{\label{tab:results}The performance of the agents trained with $ba=1$ at various risk aversions on the real market data of Figure~\ref{fig:prices}. On each scenario the p\&l and path volatility of the delta hedging strategy (last line) is compared with those of the agents (the other lines). The agents always outperform delta hedging in terms of p\&l, at the cost of an increase of the path volatility which is null or negligible for high values of $\lambda$ and it's very moderate in the other cases.}
\end{table}

\section{Conclusion}

In this paper we tackled the credit index option hedging problem with the use of reinforcement learning. As we are in a dealer market scenario, there is no market impact, thus the only way to reduce costs when trading is by optimizing the trading policy. We showed that through the use of a state of the art RL algorithm TRVO, it is possible to learn a strategy which beats the practitioner's delta hedge in terms of risk, reward and generates lower transaction costs. This result was obtained not only on data generated through a GBM, but also when generating the underlying with a heston process and using real market data. 
Interesting future works would be to consider a portfolio of options, or to complicate the financial environment by considering hybrid options.

\bibliographystyle{apacite}
\bibliography{bibliography}

\end{document}